\documentclass[useAMS,usegraphicx,usenatbib]{mn2e}

\title[SHOCK DYNAMICS IN RELATIVISTIC JETS]{SHOCK DYNAMICS IN RELATIVISTIC JETS}
\author[Cant\'o et al.]{J. Cant\'o$^{1}$, S. Lizano$^{2}$\thanks{E-mail:
s.lizano@crya.unam.mx},
M. Fern\'andez-L\'opez$^{1,3}$, R. F. Gonz\'alez$^2$, 
\newauthor and A. Hern\'andez-G\'omez$^{1,4}$\\
$^{1}$Instituto de Astronom\'\i a, UNAM, Apdo. Postal 70-264, 04510 M\'exico D. F.,
M\'exico\\
$^{2}$Centro de Radioastronom\'{\i}a y Astrof\'{\i}sica, UNAM, Apdo. Postal 3-72,
Morelia, Michoac\'an 58089, M\'exico\\
$^3$Department of Astronomy, University of Illinois, 1002 West Green Street,
Urbana, IL 61801, USA\\
$^4$Instituto de Ciencias F\'\i sicas, UNAM, Apdo. Postal 48-3, Cuernavaca,
Morelos 62210, M\'exico}

\newcommand{\be}{\begin{equation}}
\newcommand{\ee}{\end{equation}}

\newcommand{\ws}{{\rm ws}}
\newcommand{\mws}{m_{\rm ws}}
\newcommand{\gammaws}{\gamma_{\rm ws}}
\newcommand{\betaws}{\beta_{\rm ws}}
\newcommand{\T}{{\cal T}}

\begin{document}

\date{January 26, 2013}

\pagerange{\pageref{firstpage}--\pageref{lastpage}} \pubyear{2002}

\maketitle

\label{firstpage}

\begin{abstract}
We present a formalism of the dynamics of internal shocks in 
relativistic jets where the source has a time-dependent injection velocity and
mass-loss rate. The variation of the injection velocity produces a two-shock 
wave structure, the working surface, that moves along the jet. 
This new formalism takes into account the fact that 
momentum conservation is not valid for relativistic flows where 
the relativistic mass lost by radiation must be taken into account,
in contrast to the classic regime. We find analytic solutions for the
working surface velocity and radiated energy for the particular case
of a step function variability of the injection parameters. 
We model two cases: a pulse of fast material and a pulse of slow
material (with respect to the mean flow).
Applying these models to gamma ray burst light curves, 
one can determine the ratio of the Lorentz factors $\gamma_2/\gamma_1$ and the 
ratio of the mass-loss rates $\dot m_2/\dot m_1$ of the upstream and downstream flows.
As an example, we apply this model to the sources GRB 080413B and 
GRB 070318 and find  the values of these ratios. 
Assuming a Lorentz factor $\gamma_1=100$, we further estimate jet mass-loss rates 
 between
$\dot m_1 \sim 10^{-5}  - 1 \, {\rm M}_\odot \, {\rm yr}^{-1}$.
We also calculate the fraction of the injected mass lost by radiation. For
GRB 070318 this fraction is $\sim$ 7 \%. In contrast, for GRB 080413B this fraction
is larger than 50\%; in this case radiation losses clearly affect the dynamics of
the internal shocks.

\end{abstract}


\section{Introduction}

Collimated outflows (with jet-like geometry) moving at relativistic speeds
are characteristic of active galactic nuclei. It is commonly accepted that
an extragalactic jet is produced in the neighborhood of a massive black
hole in the center of an active galaxy (e.g. Rees 1984; Istomin 2010). 
These relativistic jets are subject to the development of
shock waves. Rees (1978) proposed that the observed knots in the
extragalactic jet M87 correspond to the locations of internal shocks
which arise owing to variations in the outflow velocity of a beam
generated in the nucleus. Later, Rees $\&$ M\'esz\'aros (1994)
pointed out that fluctuations of the Lorentz factor around its mean value
in a relativistic outflow, that give rise to internal shocks, can dissipate
a substantial fraction of the outflow energy into non-thermal radiation.
They proposed that this mechanism is operating in the so-called gamma-ray
bursts (GRBs).

Several authors have studied internal shocks in ultra-relativistic outflows  to explain the observed
variability of GRBs (e.g., Mochkovitch, Maitia \& Marques 1995; 
Kobayashi, Piran $\&$ Sari 1997; Daigne $\&$ Mochkovitch 1998; 2000). In these models
the flow is represented by a succession of shells with different
values of the Lorentz factor. This models reproduce
the burst profiles and their short-time scale variability.
Kobayashi et al. pointed out that variations of the relativistic flow velocity are strongly correlated with
the temporal variations observed in the GRBs. 
Daigne \& Mochkovitch (1998) studied the detailed radiation processes to calculate
the fraction of the kinetic energy dissipated in the shocks that can be emitted in the form 
of gamma rays and obtained that the total efficiency is of the order of only a few percent. 
In addition, Spada et al. (2001) proposed
that the internal shock scenario can also be used for blazars. For comprehensive
reviews of the physical processes and observations of GRBs see, e.g.,
M\'esz\'aros (2002); Piran (2004) ; and Gehrels et al. (2009).

Using mass and momentum conservation, Cant\'o et al. (2000) solved
the dynamics of internal shocks of non relativistic jets
with time dependent injection velocity and mass-loss rate. 
 Mendoza et al. (2009) used this momentum conserving formalism in the case of relativistic
jets and compared with observed light curves of GRBs assuming a sinusoidal velocity variation.
However, momentum is not conserved in relativistic
flows because radiative losses change the relativistic mass since the Lorentz factor decreases
when energy is radiated away.
In this paper we present a new formalism that describes the dynamics of internal shocks 
in a relativistic jet taking into account the momentum change by radiation. 
In particular, we study the dynamics of internal shocks for the case where the
injection velocity and mass-loss rate are both step functions of time, when
a fast wind reaches a slower flow. For this type
of variability, we find analytic solutions for the dynamical evolution
and luminosity of the shocks, which are implicit functions of time. 

The organization of the paper is as follows.  In \S 2 we discuss
the relevant relativistic equations. In \S 3 we
present the model, and investigate the dynamical evolution of the
internal WS. The luminosities predicted by our model from relativistic
jets are given in \S 4. A comparison between our analytic solutions and
 observations of extragalactic gamma ray bursts are presented
in \S 5. In \S 6 we evaluate the mass lost by radiation in the WS.
Finally, in \S 7 we summarize our conclusions.

\section[]{Relativistic equations for internal shocks}

\begin{figure}
\includegraphics[width=60mm]{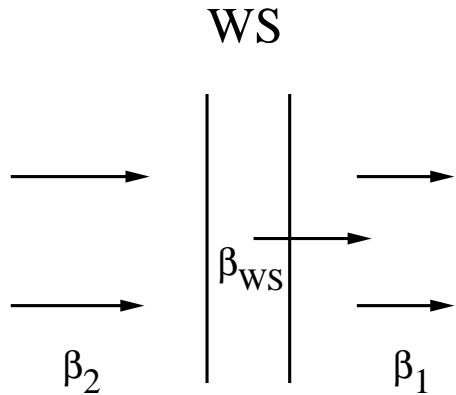}
 \caption{Schematic diagram showing a working surface formed by the interaction
between two relativistic flows. The upstream and downstream flow velocities are 
$\beta_2$ and $\beta_1$ (with $\beta_2> \beta_1$), respectively. The working
surface moves with an intermediate velocity $\beta_{ws}$.}
\label{working_surface}
\end{figure}

For a free-streaming flow, the non-dimensional velocity at a distance $x$ from the
source, at time $t$ measured at the source reference frame, is given by
\be
\beta(x,t) = \beta(0,\tau) = \frac{x}{c\left(t-\tau\right)},
\label{free_st}
\ee
where the  injection point is at $x=0$, and
$\tau$ is the time at which the flow was ejected, and, as usual, $\beta=v/c$,
where $v$ and $c$ are the speeds of of the flow and of light, respectively .  
If the flow velocity at the injection point increases, the fast material
will reach the slower material and form a working surface (WS) bounded by two
shock fronts (Raga et al. 1990). These WS structures are the so-called ``internal shocks''.
Figure \ref{working_surface} shows a schematic 
diagram of a WS formed when a fast upstream material with velocity $\beta_2$ reaches a slower
downstream material with velocity $\beta_1$. The WS moves with velocity $\betaws$ intermediate
between $\beta_1$ and $\beta_2$.
 
The internal WS forms at the distance $x_i$ from the source given by
\be
\frac{x_{i}}{c} = \left[\frac{\beta(\tau)^2}{d\beta(\tau)/d\tau} \right]_{\rm min},
\label{minimum}
\ee
where we wrote for simplicity, $\beta(\tau)= \beta(0,\tau)$, and the minimum is taken over the time interval where the velocity
increases.  The WS is formed at time $t_i$ from the
material ejected at time $\tau_i$, 
both values given by equation (\ref{minimum})\footnote{
First, $\tau_i$ is obtained by minimizing the RHS of this equation.}. Then,
one can obtain the time, $t_i$, at which the WS is formed from equation (\ref{free_st}).

Consider a relativistic jet with time dependent injection velocity, $\beta(\tau)$, and mass-loss rate, $\dot m(\tau)$.
 As discussed above, a WS is formed when fast material overtakes slow material.
This WS travels downstream the jet flow with a velocity
$\beta_\ws(t)$, where the time $t$ is measured in the source reference frame.
The slow and the fast material just entering the WS at time $t$ were ejected at times
$\tau_1$ and $\tau_2$, respectively,
 with corresponding downstream and upstream velocities 
$\beta_1= \beta(\tau_1)$ and $\beta_2= \beta(\tau_2)$, and mass-loss rates,
$\dot m_1= \dot m(\tau_1)$  and $\dot m_2= \dot m(\tau_2) $. 
The time dependence of the velocities and mass-loss rates is
given through the time dependences $\tau_1(t)$, and $\tau_2(t)$.

In Appendix A we discuss a simple example of the inelastic collision of 3 relativistic particles that radiate
energy after they collide.
This example shows that the momentum of the final particle is not conserved because its relativistic
mass changes when energy is radiated away. Therefore, we introduce below
the energy and momentum equations that
take into account the energy lost by radiation.

The total energy dissipated  by the flow interaction $E(t)$ is given by the difference between the 
total energy injected into the WS and the energy carried by the WS at the instant $t$
\footnote{For simplicity, the internal energy of the particles that enter the WS is ignored.},
\be
\frac{E(t)}{c^2} = \int_{\tau_1}^{\tau_2} \dot m(\tau) \gamma(\tau) d\tau - \mws(t) \gammaws(t) ,
\label{energy0}
\ee
where $\gammaws(t) =1/\sqrt{ 1 - \betaws(t)^2}$, and the rest mass injected into the WS is 
\be
\mws(t) = \int_{\tau_1}^{\tau_2} \dot m(\tau) d \tau .
\label{restmass}
\ee

We assume that  dissipated energy $E(t)$ is completely
radiated away (see eq. [\ref{Rad}]); none of this energy is stored in internal degrees of freedom.
Therefore, the luminosity (the radiated energy per unit time in the WS) 
is given by the time derivative $L_r(t)=dE(t)/dt$.
Then, the dynamics of the WS is described by  the energy equation,
\be
{d \over dt} \left(\mws(t) \gammaws(t) \right) = {d \over dt} \int_{\tau_1}^{\tau_2} \dot m(\tau) \gamma(\tau) d \tau - {L_r(t) \over c^2},
\label{energy_eq}
\ee
obtained from the derivative of equation (\ref{energy0}),
and the momentum equation is given by
\begin{eqnarray}
{d \over dt} \left( \mws (t) \gammaws(t) \betaws(t) \right) & =&
{d \over dt} \int_{\tau_1}^{\tau_2} \dot m(\tau) \gamma(\tau) \beta(\tau) d \tau 
\nonumber \\
& & - {L_r (t)\over c^2} \betaws(t),
\label{mom_eq}
\end{eqnarray}
where the last term in the RHS is the momentum change due to the relativistic mass lost by radiation.
Combining equations (\ref{energy_eq}) and (\ref{mom_eq}) one obtains the equation for the velocity of the WS,
\begin{eqnarray}
\mws(t) \gammaws(t) {d \betaws \over d t}  =
{d \over d t} \int_{\tau_1}^{\tau_2} \dot m(\tau) \gamma(\tau) \beta (\tau) d \tau \nonumber \\
  -\betaws (t){d \over d t} \int_{\tau_1}^{\tau_2} \dot m(\tau) \gamma(\tau) d \tau 
\nonumber  \\
=  \dot m(\tau_2) \gamma(\tau_2)\left[ \beta(\tau_2) -\betaws(t)\right] {d \tau_2 \over d t} \nonumber \\ 
  - \dot m(\tau_1) \gamma(\tau_1)\left[ \beta(\tau_1) -\betaws(t)\right] {d\tau_1 \over d t},
\label{princ_eq}
\end{eqnarray}
that does not depend explicitly on the radiated energy, $L_r$. 

Given a time variability of the functions $\beta(\tau)$ and $\dot m(\tau)$,
one can find the WS velocity, $\betaws(t)$, from  equation (\ref{princ_eq}) together with equation (\ref{free_st}), the latter giving
the relation between $\tau_{1,2}$ and $t$.
We will show below how these equations can be solved
analytically for the case of step function variability of the injection parameters.

\section{Step function variability of the injection parameters}

\begin{figure}
\includegraphics[width=84mm]{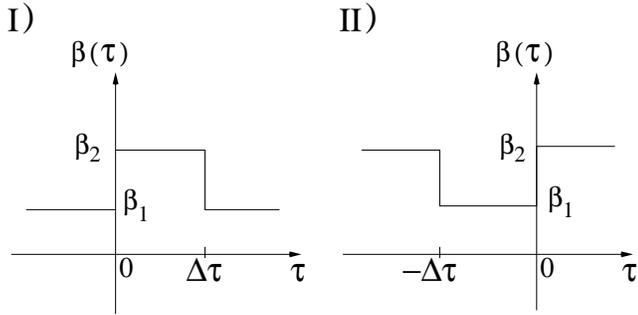}
\caption{Injection velocity $\beta$ as function of time $\tau$: I) The initial velocity
$\beta_1$ suddenly increases to $\beta_2$ at a time $\tau= 0$, for a finite time interval,
$\Delta \tau$, and then instantly returns back to its original value.
II) The initial velocity $\beta_2$ instantly decreases to $\beta_1$ at a time
$\tau= -\Delta \tau$, and at $\tau= 0$,  the faster flow starts to be injected again.}
\label{injection_vel}
\end{figure}

Consider a step function variability for the injection velocity and mass-loss rate 
such that, for $\tau < 0$, a slow flow has $\beta = \beta_1$,  and $\dot m = \dot m_1$; 
and for $\tau \ge 0$, a fast flow has $\beta=\beta_2$ and $\dot m = \dot m_2$.   
Figure \ref{injection_vel}
shows two cases: I) when fast material is injected at $\tau > 0$ for a
finite time, $\Delta \tau$; and II) when slow material is injected at $\tau < 0$,
for a finite time interval, $\Delta \tau$.  
In both cases, when the fast flow reaches the slow flow,  the WS
is formed instantaneously at the injection point:
$x=0$, $t=0$, and $\tau=0$.

As we will show below, the
dynamical evolution of the WS goes through 2 stages. In the first stage,
the WS is fed by both the slow and fast flows and  it moves at a constant velocity,
intermediate between the fast and slow speeds. The second stage begins when one of the flows
has been completely  incorporated into the WS; then, the WS accelerates or decelerates depending 
on which flow (fast or slow) continues to feed the remaining shock. Asymptotically in time, the speed
of the WS in the second stage tends to the velocity of the
remaining flow. 

\begin{figure}
\includegraphics[width=54mm]{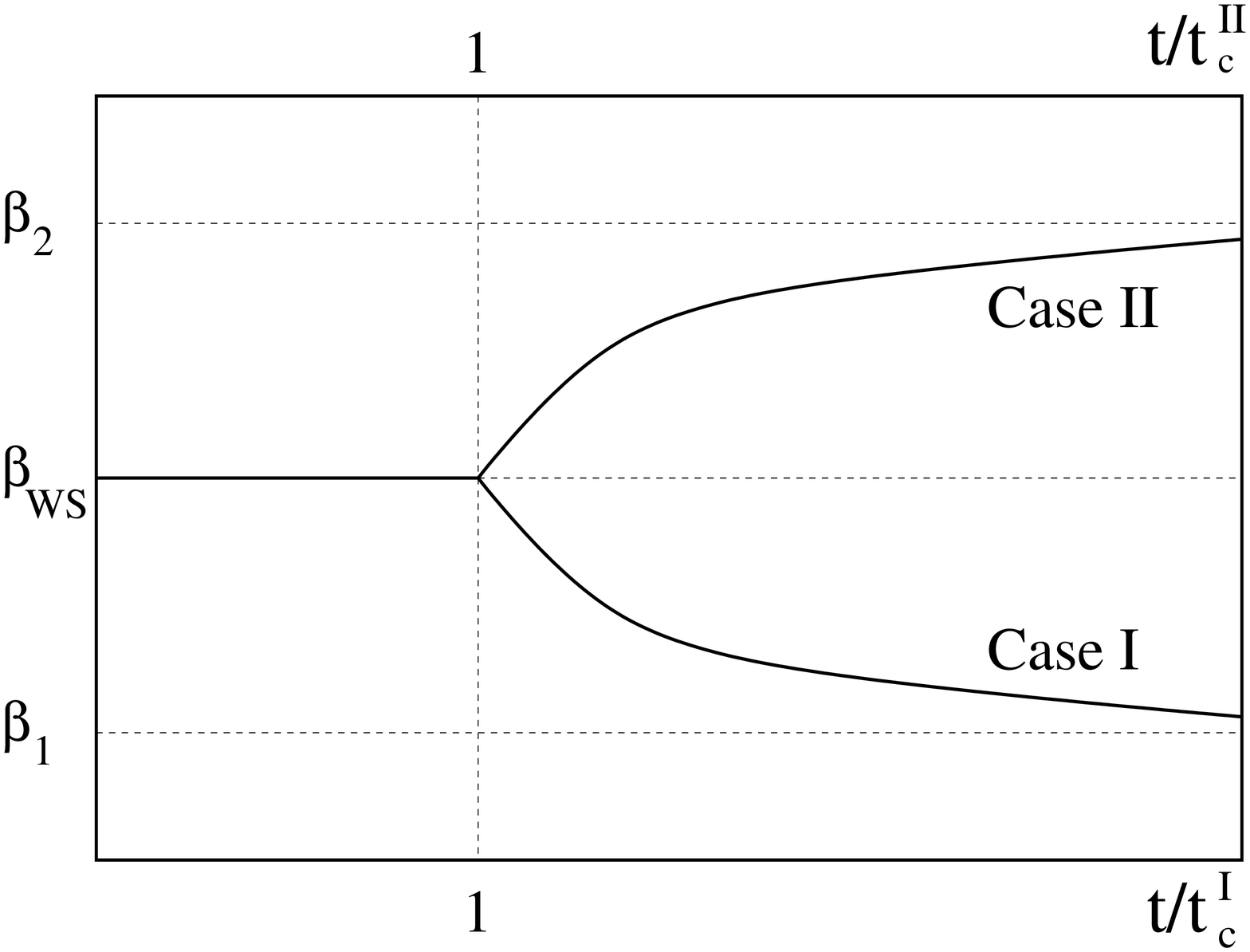}
\caption{ Qualitative behavior of the WS velocity as a function of normalized time $t/t_c^{I,II}$, for Case I and Case II,
respectively. See text for description of this figure.}
\label{vws_t}
\end{figure}

Figure \ref{vws_t} shows the
qualitative behavior of the WS velocity as a function of normalized time for Case I (pulse of fast material) and Case II (pulse of slow material).
In Case I, the time is normalized to the critical time $t_{c}^I$ (eq. [\ref{tcI}]).
In Case II, the time is normalized to the critical time $t_{c}^{II}$ (eq. [\ref{tcII}]).
In both cases, the constant velocity phase ends when $t/t_c^{I,II}=1$.

Consider the material at time $t$  that enters the WS, located at the position  $x_{\rm ws}(t)$, 
through both shocks. The slow and fast material were
ejected at times $\tau_1$ and $\tau_2$, respectively, such that, according to equation (\ref{free_st}) 
\be
\tilde \tau_{1,2} = \tilde t -\frac{\tilde x_{\rm ws}}{ \beta_{1,2}}  \quad {\rm and} 
\quad {d \tilde \tau_{1,2} \over d \tilde t} = 
1 - {\betaws \over \beta_{1,2}},
\label{tau12}
\ee
where  $\tilde \tau_{1,2}=\tau_{1,2}/\Delta \tau$, $\tilde t = t / \Delta \tau$, and $\tilde x_{\rm ws} = x_{\rm ws} / (c \Delta \tau)$.

Now, for a step function variability,
the rest mass of the WS, given by equation (\ref{restmass}), is
\begin{eqnarray}
m_{\rm ws}(t)  & = &\int_{\tau_1}^{0} \dot m_1(\tau) d \tau +
                \int_{0}^{\tau_2}  \dot m_2(\tau) d \tau \nonumber  \\
& = & - \dot m_1 \tau_1(t) + \dot m_2  \tau_2(t) .
\label{mphase1}
\end{eqnarray}

Note that $\tau_1 < 0$ and $\tau_2 > 0$, and for now on, for simplicity, we will drop the $t$ dependence
of the functions.

The velocity of the WS is obtained from equation (\ref{princ_eq}) 
\be
\mws \gammaws {d \betaws \over d t}  =
\dot m_2 \gamma_2  {\left(\beta_2- \betaws  \right)^2\over \beta_2} -  \dot m_1 \gamma_1{ \left( \beta_1  - \betaws \right)^2 \over \beta_1} ,
\ee
where we used equation  (\ref{tau12}). This equation has the constant velocity, $\beta_{\rm ws0}$,
solution \footnote{This solution exists because at $t=0$, the initial condition $\beta_{\rm ws0}$ makes the RHS
of the equation equal zero, since $m_\ws(0) =0$.},
such that $d \beta_{\rm ws0}/dt=0$, then, 

\be
\lambda \left( {\beta_2 -\beta_{\rm ws0}}\right) = \pm \left({\beta_{\rm ws0} - \beta_1} \right),
\label{solpm}
\ee
with
\be
\lambda = \sqrt{\dot m_2 \gamma_2 \beta_1 \over \dot m_1 \gamma_1 \beta_2} = \sqrt{{b r \over a}},
\label{lambda}
\ee
where we defined the  velocity ratio $a=\beta_2/\beta_1$, the mass-loss rate ratio $b=\dot m_2/\dot m_1$, and
gamma ratio $r= \gamma_2/\gamma_1$.
The correct solution corresponds to the $+$ sign
\footnote{The $-$ sign in equation (\ref{solpm}) is unphysical because, in this solution, 
$\beta_{ws0} < \beta_1$, that implies that there is no downstream shock.},
with the ordering $\beta_1 < \betaws < \beta_2$, where the constant
WS velocity  is given by
\begin{eqnarray}
\beta_{ws0}= {\lambda \beta_2 + \beta_1 \over \lambda + 1}, 
\quad {\rm and}
\nonumber
\end{eqnarray}
\be
\gamma_{ws0}=
{\lambda+1 \over 
\sqrt{ \left( {\lambda/ \gamma_2} \right)^2  + 2 \lambda\left(1-\beta_1\beta_2\right) +
1/\gamma_1^2  } } .
\label{beta_first}
\ee
This velocity corresponds to the first stage of the evolution of the WS, when it is fed (and bounded) by two shocks.

Now we will discuss the evolution of the WS in the second stage of the
two different cases I and II shown in Figure \ref{injection_vel}. 
Even though the formalism is the same, the resulting equations are different for each case. 
Thus, for clarity, we separate them in the next two subsections.

\subsection{Case I: Decelerating WS}

The constant velocity phase ends when $\tau_2 = \Delta \tau$, i.e., when the fast material is completely incorporated into 
the WS. This happens at a  critical time obtained by substituting the position of the WS, 
$\tilde x_{ws0}=\beta_{ws0} \tilde t$,  into equation (\ref{tau12}),
\be
\tilde  t_{\rm c}^{I}  =  {a \left( \lambda + 1 \right) \over a-1},
\label{tcI}
\ee
which corresponds to ejection time
\be
\tilde \tau_{1, \rm c} =  
\tilde t_{\rm c}^{I}\left( 1 - {\beta_{\rm ws0} \over  \beta_1 } \right) = -a \lambda .
\label{tau1c}
 \ee

In this second stage, $\tilde t > \tilde t_{\rm c}^{I}$, the first term of
equation (\ref{princ_eq}) is $0$ because $d \tau_2 / d t =0$. Furthermore, substituting $\tau_2=\Delta \tau$ 
in equation (\ref{mphase1}), 
the rest mass of the WS is
\be
\mws = \dot m_2 \Delta \tau - \dot m_1 \tau_1 = \dot m_1 \Delta \tau \left (b - \tilde \tau_1 \right).
\label{mwsI}
\ee
Collecting these results
we write the WS velocity in equation (\ref{princ_eq}) as function of $\tilde \tau_1$ 
as
\be
{d  \betaws \over d \tilde \tau_1}  =
{ \gamma_1 \left(  \betaws - \beta_1\right) \over \gammaws \left( b - \tilde \tau_1 \right) }.
\label{dbwsdtau1}
\ee
For $\betaws \ne \beta_1$, this equation \footnote{Equation \ref{dbwsdtau1} has a constant velocity solution, 
 $\betaws=\beta_1$, that is trivial because there is no shock.}
can be integrated by separation of
variables as
\be
{2 \left(\gamma_1-\betaws \beta_1\gamma_1 + 1/\gammaws  \right)
\over \left(\betaws-\beta_1\right)  } = C^{I} \left( b -\tilde \tau_1 \right),
\label{betaws_implicit}
\ee
that is a quadratic equation for $\betaws (\tilde \tau_1)$, and has the solution
\begin{eqnarray}
\betaws (\tilde \tau_1)={\beta_1 {\T}^2 + 4 {\T} + 4 \beta_1 \over \T^2 + 4 \beta_1 \T + 4} \quad {\rm and}
\nonumber
\end{eqnarray}
\be
\gamma_{\rm ws}(\tilde \tau_1) = \gamma_1 \left( {\T^2 + 4 \beta_1 \T + 4 \over \T^2 - 4 } \right),
\label{betawstau1}
\ee
where we defined the time function
\be
\T (\tilde \tau_1)= {C^{I} (b - \tilde \tau_1) \over \gamma_1},
\label{calT1}
\ee
that is an increasing function of $|\tilde \tau_1|$. From equation (\ref{betaws_implicit}) one can show that
$\T(\tilde \tau_{1}) > 2$. 
The constant $C^{I}$ is obtained by matching the solution $\betaws(\tilde \tau_{\rm 1, c} )=\beta_{ws0}$,
 where $\tilde \tau_{1,\rm c}$ is given by equation (\ref{tau1c}); thus, from
 equation (\ref{betaws_implicit}) one gets
\be
C^{I} = 
{2 \left(\gamma_1-\beta_{\rm ws0} \beta_1 \gamma_1+ 1/\gamma_{\rm ws0} \right)
\over \left(b + a \lambda \right) \left(\beta_{\rm ws0}-\beta_1\right) }.
\ee
To find a relation between $\tilde \tau_1$ and $\tilde t$,  we take 
the derivative of the time function $\T$ (eq. [\ref{calT1}]),
\be
{d \T \over d \tilde t} = -{C^{I} \over \gamma_1} {d \tilde \tau_1 \over d \tilde t}
= {4 C^{I} \over \beta_1 \gamma_1^3} {  \T \over 
 \left( \T^2 + 4 \beta_1 \T + 4 \right) },
 \label{dcalTdt}
\ee
 where we used equations (\ref{tau12}) and (\ref{betawstau1}).
 Again, by separation of variables, this equation can be integrated as
 \be
 {\T^2 \over 2} + 4 \beta_1 \T + 4 \ln{\T} = {4 C^{I} \over \beta_1 \gamma_1^3} \tilde t + D^{I},
 \label{calT1_t}
 \ee
 where the constant $D^{I}$ is obtained evaluating this expression at  $\tilde t_{\rm c}^{I}$ and $\tilde \tau_{1, \rm c}$, 
 and is given by,
 \be
 D^{I} = {\T_{\rm c}^2 \over 2}  + 4 \beta_1 \T_{\rm c}
 + 4 \ln{\T_{\rm c}} - {4 C^{I}  \over \beta_1 \gamma_1^3 } 
 { a \left(\lambda +1\right)  \over \left(a-1\right)} ,
 \ee
where the critical time function is
\be
\T_{\rm c} = \T(\tilde \tau_{1,\rm c})= C^{I} {(b + a \lambda)\over \gamma_1}.
\label{calTc1}
\ee

Therefore, given $\tilde \tau_1$ one
can evaluate the time function $\T$ to obtain $\betaws(\tilde \tau_1)$ 
from equations (\ref{betawstau1}) and (\ref{calT1}), and 
one can obtain the WS velocity as a function of time $\tilde t$, $\betaws(\tilde t)$,
 in a tabular form, using equation (\ref{calT1_t}). 
Finally, the position of the WS, $x_{\rm ws}(\tilde t)$, is given by equation (\ref{tau12}).

\subsection{Case II: Accelerating WS}

 In this case, the constant velocity phase ends when $\tau_1 = -\Delta \tau$, i.e., when the slow
material is completely incorporated into 
the WS. 
From equation (\ref{tau12}), the critical time is
\be
\tilde  t_{\rm c}^{II} =  { \left( \lambda + 1 \right) \over \lambda\left( a-1\right)},
\label{tcII}
\ee
corresponding to the ejection time
\be
\tilde \tau_{2, \rm c} =  
\tilde t_{\rm c}^{II}\left( 1 - {\beta_{\rm ws0} \over  \beta_2 } \right) = {1\over a \lambda }.
\label{tau2crit}
 \ee 

 For $\tilde t > \tilde t_{\rm c}^{II}$, one has $d \tau_1 / d t =0$. Also,
from equation (\ref{mphase1}), the rest mass is 
\be
\mws = \dot m_2 \tau_2 + \dot m_1 \Delta \tau = \dot m_1 \Delta \tau \left (b  \tilde \tau_2 +1 \right).
\label{mwsII}
\ee
Then, we write the WS velocity in equation (\ref{princ_eq}) as function of $\tilde \tau_2$ as
\be
{d  \betaws \over d \tilde \tau_2}  =
{ \gamma_2 b  \left( \beta_2- \betaws \right) \over \gammaws \left( b \tilde \tau_2+1 \right) }.
\label{dbwsdtau2}
\ee

For $\betaws \ne \beta_2$, this equation \footnote{Equation \ref{dbwsdtau2} has a constant velocity solution, 
 $\betaws=\beta_2$, which is trivial because there is no shock.} can be integrated by separation of
variables as
\be
{2 \left(\gamma_2-\betaws \beta_2\gamma_2 + 1/\gammaws  \right)
\over \left(\beta_2-\betaws\right)  } = C^{II} \left( b \tilde \tau_2+1 \right),
\label{betawsII_implicit}
\ee
which is a quadratic equation for $\betaws (\tilde \tau_2)$, and has the solution
\begin{eqnarray}
\betaws (\tilde \tau_2)={\beta_2 {\T}^2 - 4 {\T} + 4 \beta_2 \over \T^2 - 4 \beta_2 \T + 4}\,, \quad {\rm and}
\nonumber
\end{eqnarray}
\be
\gamma_{\rm ws}(\tilde \tau_2) = \gamma_2 \left( {\T^2 - 4 \beta_2 \T + 4 \over \T^2 - 4 } \right),
\label{betawstau2}
\ee
where the time function is
\be
\T (\tilde \tau_2)= {C^{II} (b  \tilde \tau_2+1) \over \gamma_2},
\label{calT2}
\ee 
that is an increasing function of $\tilde \tau_2$. As in the previous case, 
one can show that
$\T(\tilde \tau_{2}) > 2$. 
The constant $C^{II}$ is obtained by matching the solution $\betaws(\tilde \tau_{\rm 2, c} )=\beta_{ws0}$,
 where $\tilde \tau_{2,\rm c}$ is given by equation (\ref{tau2crit}).
Substituting these
 results into equation (\ref{betawsII_implicit}) one gets
\be
C^{II} = 
{2 a \lambda \left(\gamma_2-\beta_{\rm ws0} \beta_2 \gamma_2+ 1/\gamma_{\rm ws0} \right)
\over \left(b + a \lambda \right) \left(\beta_2-\beta_{\rm ws0}\right) }.
\ee
We take 
the derivative of the time function  $\T$ (eq. [\ref{calT2}]),
\be
{d \T \over d \tilde t} = {C^{II} b\over \gamma_2} {d \tilde \tau_2 \over d \tilde t}
= {4 C^{II} \over \beta_2 \gamma_2^3} {  \T \over 
 \left( \T^2 - 4 \beta_2 \T + 4 \right) },
 \label{dcalT2dt}
\ee
where we used equations (\ref{tau12}) and (\ref{betawstau2}).
By separation of variables, this equation can be integrated as
 \be
 {\T^2 \over 2} - 4 \beta_2 \T + 4 \ln{\T} = {4 C^{II} \over \beta_2 \gamma_2^3} \tilde t + D^{II},
 \label{calT2_t}
 \ee
 where the constant $D^{II}$ is obtained evaluating this expression at  $\tilde t_{\rm c}^{II}$ and $\tilde \tau_{2, \rm c}$, 
 and is given by,
 \be
 D^{II} = {\T_{\rm c}^2 \over 2}  - 4 \beta_2 \T_{\rm c}
 + 4 \ln{\T_{\rm c}} - {4 C^{II}  \over \beta_2 \gamma_2^3 } 
 {  \left(\lambda +1\right)  \over \lambda \left(a-1\right)} ,
 \ee
where the critical time function is 
\be
\T_{\rm c} = \T(\tilde \tau_{2,\rm c})= C^{II} {(b + a \lambda)\over a \lambda \gamma_2}. 
\label{calTcrit2}
\ee
 Using equations (\ref{tau12}), (\ref{betawstau2}), (\ref{calT2}), and (\ref{calT2_t}) 
one can proceed as
in Case I to obtain $\betaws(\tilde t)$ and  $x_{\rm ws}(\tilde t)$ as functions of $t$ in a tabular form.

\begin{figure}
\includegraphics[width=74mm]{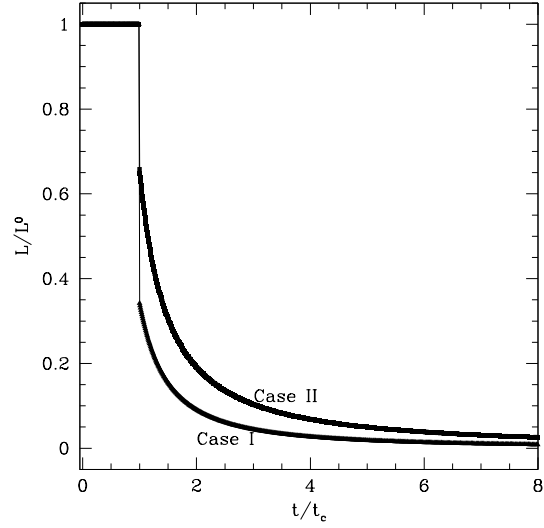}
\caption{Normalized luminosity $ L/ L^0$, as function of the normalized time $t/t_{c}$ for 
Case I and Case II. For these models, we assumed
$\gamma_1=100$ and $\gamma_2=200$. We also assumed $\lambda =1$,
thus, the critical times are the same for both cases. }
\label{L2cases}
\end{figure}

\section{Luminosities}

Using a step function variability of the injection parameters in equation
(\ref{energy_eq}), 
the luminosity $L_r$ of the WS is given by 

\begin{eqnarray}
{L_r(\tilde t) \over c^2} 
&=&  \dot m_2 \gamma_2{d \tilde \tau_2 \over d \tilde t} 
-\dot m_1 \gamma_1{d \tilde \tau_1 \over d \tilde t} \nonumber \\
&& + \left( \dot m_1 {d \tilde \tau_1 \over d\tilde  t}  - 
 \dot m_2 {d \tilde \tau_2 \over d \tilde t}\right) \gammaws
- {m_\ws \over \Delta \tau} {d \gammaws \over d \tilde  t} \nonumber \\
& = &
\dot m_2 \left ( \gamma_2 -\gammaws \right) {d \tilde \tau_2 \over d \tilde t}
+ \dot m_1 \left( \gammaws - \gamma_1 \right) {d \tilde \tau_1 \over d \tilde t} \nonumber \\
&&  -  {m_{\ws} \over \Delta \tau} {d \gammaws \over d \tilde t}.
\label{lum}
\end{eqnarray}
where we have used equation (\ref{mphase1}) for the rest mass $m_\ws$. 

Only a fraction of the energy radiated in internal shocks will be emitted as gamma rays; this fraction 
is low, $\epsilon \sim 0.01$ (e.g., Daigne \& Mochkovitch 1998).
Here we will assume that a constant fraction $\epsilon$ of the luminosity will go into gamma ray
radiation, $L_{GRB} = \epsilon L_r(t)$, and express this GRB luminosity 
in non dimensional form as
\begin{eqnarray}
\tilde L &= &\frac{L_{GRB}}{ \epsilon c^2 \gamma_1 \dot m_1} \nonumber \\
&= & a \lambda^2
 \frac {\left( \gamma_2 -\gamma_\ws \right ) }{\gamma_2} \frac{ d \tilde \tau_2}{d \tilde t} 
 + \frac{ \left( \gamma_\ws -\gamma_1 \right)}{\gamma_1} \frac{d  \tilde \tau_1}{d \tilde t}\nonumber \\
& & - {m_\ws \over  \dot m_1 \gamma_1 \Delta \tau} \frac{d \gamma_\ws} {d \tilde t}.
\label{LGRBnondim}
\end{eqnarray}
In the constant velocity phase, $\tilde t < \tilde t_{c}^{I,II}$, the luminosity is
given by
\begin{eqnarray}
\tilde L^{0} & = & 
 \lambda^2
 \frac {\left( \gamma_2 -\gamma_{ws0} \right ) }{\gamma_2} \left( a - {\beta_{ws0}\over \beta_1}  \right)
\nonumber \\
& & - \frac{ \left( \gamma_{ws0} -\gamma_1 \right)}{\gamma_1}  \left( { \beta_{ws0} \over \beta_1} -1 \right) 
\nonumber \\
& = & \left( { \beta_{ws0} \over \beta_1} -1 \right)
\left[
\lambda
 \frac {\left( \gamma_2 -\gamma_{ws0} \right ) }{\gamma_2} 
 - \frac{ \left( \gamma_{ws0} -\gamma_1 \right)}{\gamma_1} \right] ,
\label{L_0}
\end{eqnarray}
where we substituted  $d \tilde \tau_1 / d \tilde t = 1 - \beta_{\rm ws}/ \beta_1$
and $d \tilde \tau_2 / d \tilde t = 1 - \beta_{\rm ws}/ \beta_2$ from equation (\ref{tau12}).

As mentioned in \S 3.1, in Case I, once all the fast material has been been
completely incorporated into the working surface, it is decelerated
by the slow downstream flow. This second stage starts at a time
$\tilde t_{c}^I$ given by equation (\ref{tcI}).  For this case, 
on has $d \tilde \tau_2 / d \tilde t = 0$ in equation (\ref{lum}); thus, one can write the luminosity as,
\begin{eqnarray}
\tilde L^I & = &- {{\left( \gammaws - \gamma_1 \right)}\over{\gamma_1}}  \left( { \beta_{ws0} \over \beta_1} -1 \right)
 - {{m_{\rm ws}}\over{ \dot m_1 \gamma_1  \Delta \tau}}
 {d \gammaws \over d \tilde t}  \nonumber \\
 & = &  {{(\beta_{\rm ws} - \beta_1)}\over{\beta_1}}\, \left[
\beta_{\rm ws}\,(\beta_{\rm ws} - \beta_1)\,\gamma_{\rm ws}^2
- {{(\gamma_{\rm ws} - \gamma_1)}\over{\gamma_1}}
\right] .
\label{L_I}
\end{eqnarray}
where we used that $d \gammaws/ d \tilde t = \beta_{\rm ws}
\gamma_{\rm ws}^3 d \beta_{\rm ws}/ d \tilde t$, and, from equation (\ref{princ_eq}),
\begin{eqnarray}
{d \betaws \over d \tilde t} =
 - {{\dot m_1 \gamma_1 \Delta \tau} \over {\mws \gammaws}} 
\, {{(\beta_1 - \beta_{\rm ws})^2} \over {\beta_1}}  .
\label{eqmot}
\end{eqnarray}

On the other hand, in Case II, once the slow material has been completely incorporated
at $\tilde t_{c}^{II}$ given by equation (\ref{tcII}), $d \tilde \tau_1 / d \tilde t = 0$, and the luminosity is given by
\begin{eqnarray}
\tilde L^{II} & = &
\lambda^2 \frac {\left( \gamma_2 -\gamma_{ws0} \right ) }{\gamma_2} \left( a - {\beta_{ws0}\over \beta_1}  \right)
 - {{m_{\rm ws}}\over{\dot m_1 \gamma_1 \Delta \tau}}
 {d \gammaws \over d \tilde t} \nonumber \\
& =&  \lambda^2 a  {\left (\beta_2 -\betaws\right) \over \beta_2}\, \times  \nonumber \\
&& \left[ -\betaws \left( \beta_2 - \betaws \right) \gamma_\ws^2 +  {\left( \gamma_2 - \gamma_\ws \right) \over \gamma_2}
 \right] ,
 \label{L_II}
\end{eqnarray}
where, from equation (\ref{princ_eq}),
\be
{d \betaws \over d \tilde t} =
  {\dot m_2 \gamma_2 \Delta \tau \over \mws \gammaws} 
\, {\left(\beta_2 - \beta_{\rm ws}\right)^2 \over \beta_2}\,.
\ee

\begin{figure*}
\includegraphics[width=185mm]{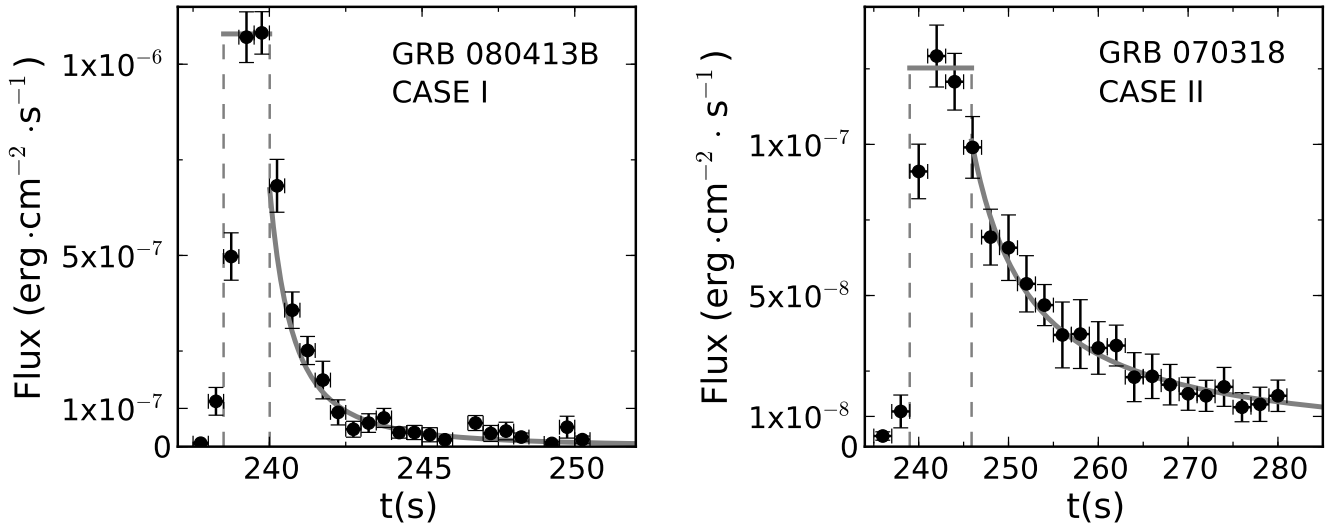}
\caption{ Left panel: GRB 080413B source with a decelerating WS model (thick solid line). 
Right panel: GRB 070318 source with an accelerating WS model (thick solid line). The thin lines
in each panel show the duration of the constant velocity phase. 
The observational data of both sources was taken from Mendoza et al. (2009).}
\label{GRB05111}
\end{figure*}

In the ultra-relativistic (UR) limit, $\gamma \gg 1$, the expressions for the luminosities are simplified as shown in
Appendix B.
In the following section we will apply these equations for the GRB luminosities to describe the light curves of two observed sources.

\section{Predicted Fluxes}

Figure \ref{L2cases} shows the model luminosity for Case I and Case II, 
normalized to the luminosity in the constant velocity phase, $ L^{I,II} /  L^0$, as 
a function of normalized time, $t/t^{I,II}_{c}$. The models
have $\gamma_1 =100$, $\gamma_2 = 200$, that correspond to the UR limit. 
Also, we choose $\lambda \sim (b r)^{1/2}= 1$, that implies a constant energy injection rate
 $\dot m_1 \gamma_1 = \dot m_2 \gamma_2$. In this case,   the critical times
(eqs. [\ref{tcI}] and [\ref{tcII}]) are equal. 
For this reason, we drop the superscripts in the following discussion.
As discussed above, for $t  < t_{c} $, the WS is bounded by 2 shocks and moves at a constant speed. 
Then, the relative
velocity between the incorporated material and the WS and, therefore,  the luminosity are constant.
 For $t \ge t_{c} $,  in both cases one shock disappears and
the relative velocities between the WS and the new material that is incorporated decreases with time;
therefore, the luminosity, diminishes with time.

Appendix \ref{alpha_fit} and \ref{fund_par} show that,  in the UR approximation,
one can fit a power-law  to the wing of the GRB light curve,
and obtain the critical time function $\T_c$, the gamma ratio, $r=\gamma_2/\gamma_1$, 
and the mass-loss rate ratio, $b= \dot m_2/\dot m_1$. These quantities can be obtained without any further assumptions.
As an example, Figure \ref{GRB05111} shows a model fit to the observed sources GRB 08413B and GRB 070318. 
 The fluxes were taken from Mendoza et al. (2009) that correspond to  observations between
15 and 150 keV with the Burst Alert Telescope on board the SWIFT satellite. 
We first fit the observed flux density $F$ directly and later discuss its relation to the distance and luminosity of
the GRB.
We follow the procedure described in Appendix \ref{fund_par}, and  choose the value of the 
 luminosity (or flux density) in the constant velocity phase, the time $t_0$ of the beginning of the velocity pulse,
 and time $t_e = t_c + t_0$ of the end of the constant velocity phase,
 where $t_c$ is the critical time of the model. We also choose the flux at the constant 
 velocity phase, $F^0$.
The wing light curve is then fit by a power-law
$F = B(t-t_0)^{\alpha}$. As discussed in the Appendix C, the value of the slope $\alpha$, 
determines which case (I or II) applies.  

Table 1 shows the model parameters that fit the light curves of both GRBs. The name of the GRB is indicated in column 1;
the applied model (Case I or Case II) is shown in column 2; column 3 shows the flux in the constant velocity phase;
column 4 and column 5 show the time of the beginning of the pulse, $t_0$, and the time of the end of the 
constant velocity phase, $t_e$; column 6 and column 7 show the coefficient, $B$ and 
the exponent $\alpha$ of the power-law fit of the GRB light curve wing; column 8 gives the inferred 
value of the critical time function, $\T_c$; column 9 gives the 
jump of the flux (or luminosity) at the end of the constant velocity phase,
 $J(t_e)$ defined by equations (\ref{LURI0}) and (\ref{LURII0}); 
 finally, column 10 and column 11 give the inferred values of the 
gamma ratio, $r$ and the mass-loss rate ratio, $b$, obtained directly form the fit to the wind of the 
light curve. 

In our model, the WS approaches the observer at relativistic speeds.
As discussed in Appendix \ref{Dopp_boost},  the observed bolometric flux density of an approaching
relativistic jet is increased with respect to an emitter at rest by a factor 
$\delta_{\rm ws}^4$, where $\delta_{\rm ws}=1/[\gammaws (1-\betaws\cos\theta)]$ and 
$\theta$ is the angle between the direction of the relativistic jet and the observer.
Lind \& Blandford (1985) obtained an amplification by a factor of $\delta_{\rm ws}^3$ between the 
observed flux at frequency $\nu$ and the emitted luminosity at frequency $\nu^\prime=\nu/\delta_{\rm ws}$. 
An extra factor of $\delta_{\rm ws}$ is obtained when one integrates the frequency
to get the observed bolometric flux  (or the observed flux in a frequency range) in terms of the emitted bolometric luminosity.
Thus, in our model the observed gamma ray flux in the constant velocity phase is given by
$F^0_{GRB} =  \delta_{\rm ws0}^4 \tilde L^0_{UR} \epsilon c^2 \gamma_1 \dot m_1/ 4 \pi D^2$, 
where $\tilde L^0_{UR}$ is the normalized luminosity  in equation (\ref{LUR0}).

In order to solve for the mass-loss rate, we assume that the relativistic jet is seen almost along the jet axis, i.e.,
$\cos\theta \sim 1$, thus, $\delta_{\rm ws0} \sim 2 \gamma_{\rm ws0}$.
Also, assuming $\gamma_1=100$, both $\tilde L^0_{UR}$ and the Lorentz factor $\gamma_{\rm ws0}$ can be obtained 
from the model parameters in Table 1. In particular, we obtain large Lorentz factors for the WS,
$\gamma_{\rm ws0} \sim 802$ for GRB 080413B, and $\gamma_{\rm ws0} \sim 120$ for GRB 070318 from equation (\ref{gammaws0}).
Furthermore, we estimate the distance to the GRBs from their redshift.
The source GRB 070318 has an estimated redshift $z=0.836$ (Jaunsen et al. 2007) and 
GRB 080413B has a redshift $z=1.1$ (Vreeswijk et al. 2008). Assuming a dark energy density 
$\Omega_\Lambda=0.7$, a matter density
$\Omega_m = 0.3$, and a Hubble constant $H=70 \,  {\rm km \, s^{-1} Mpc^{-1}}$, 
the luminosity distance of GRB 070318 is $D=5300$ Mpc and of GRB 080413B is $D = 7440$ Mpc.
With all these ingredients, we can now solve
for the mass-loss rate, and obtain
$\dot m_1 \sim 9.7 \times 10^{-6} {\rm M}_\odot \, {\rm yr}^{-1}$ for GRB 080413B,
 and $\dot m_1 \sim 1.4 \,  {\rm M}_\odot\, {\rm yr }^{-1}$ for GRB 070318.
Because the emitting WS moves with relativistic speed, 
 the jet mass-loss rates $\dot m_1$ required to produce the gamma ray flux observed at the Earth
are much smaller than those obtained, for example, in nucleosynthesis models of wind-driven supernovae 
 and collapsar models (e.g., Fujimoto et al. 2008; Maeda \& Toming 2009).
 
\begin{table*}
 \centering
 \begin{minipage}{165mm}
  \caption{Model Parameters}
  \begin{tabular}{@{}lccccccccrr@{}}
\hline
\hline
GRB & Case & $F_{GRB}^0$ \footnote{Flux density in the constant velocity phase,
$F_{GRB}^0 =  \delta^4 \tilde L^0_{UR} \epsilon c^2 \gamma_1 \dot m_1/ 4 \pi D^2$.}
& $t_0$ & $t_e$ & $B \pm \Delta B$ & $\alpha\pm \Delta
 \alpha$ & $\T_c$ & $J$ & $r$ & $b$ \\
   & & ${\rm erg \, cm^{-2} s^{-1}}$ & ${\rm (s)}$ & ${\rm (s)}$ & ($10^{-7} {\rm erg
 \, cm^{-2} s^{-1-\alpha}})$ & & & & ($\gamma_2\over \gamma_1$) & ($\dot m_2 \over
 \dot m_1$) \\
\hline
080413B & I & $1.08 \times 10^{-6}$  & 238.50 &  240.01 & $15.49 \pm 3.00 $ & $-2.03 \pm 0.30$ &2.57 & 0.38 & 22.49 & 233.73 \\
070318  & II & $1.25 \times 10^{-7}$ & 239.00 &  245.90 & $8.08  \pm  0.50$  & $-1.08  \pm 0.05 $ &5.35 & 0.20 & 2.64 & 0.12 \\ 
\hline
\end{tabular}
\end{minipage}
\end{table*}

Note that in the model the time is measured in the frame of reference of the jet source, 
where the evolution timescales for the WS are of the order of $\sim 10^{5}-10^{6}$s. 
Instead,  the time of the observations is measured at the observer's frame of reference 
and is only of the order of tens of seconds if the relativistic jet is seen almost along the 
jet axis. This happens because the arrival 
time is $\Delta t_{a} = \Delta t_{em}/(\gammaws \delta_{\rm ws})$, where 
$\Delta t_{em}$ is the emission time at the jet source frame
(e.g., Daigne \& Mochkovitch 1998). 
Also, our models are not meant to explain the shape variation of the 
bursts with spectral band (e.g., Norris et al. 1996), 
which would depend on fraction of enegy radiated 
in gamma rays, $\epsilon$. In fact, one expects that $\epsilon$ will depend on energy and time, as the WS 
decelerates and the energy is radiated at lower spectral bands.

Finally, the model uses the simplest velocity 
variation which allows an analytic solution: it assumes 
an instantaneous jump in the velocity of the 
injected material (step function), thus, it makes the simplification that the luminosity instantly 
achieves the maximum value. A more realistic situation would be a gradual increase in the velocity of the injected material 
which would produce a gradual growth in the luminosity.  
Here we show that the decay of two GRB light curves
can be fitted by the emission of an decelerated or accelerated WS 
given by these very simple models.
Although, this choice is intended to illustrate the formalism it is also true that the analytic 
solutions allow an exploration of parameter space and give us an
understanding of the dependance of the GRB emission on important physical 
parameters like the gamma ratio $r$ and the mass-loss rate ratio $b$.

\section{Fraction of the injected mass lost by radiation in the WS}

In this section, we evaluate the fraction of the mass lost by radiation
with respect to the mass  injected in the WS, $\Delta m$.
From equation (\ref{energy_eq}) the total mass lost by radiation  is
\be
\int_0^t{L_r(t)  c^{-2}} dt = \int_{\tau_1}^{\tau_2} \dot m(\tau) \gamma(\tau) d \tau -
\mws(t) \gammaws(t).
\label{mass_lost}
\ee
In the RHS of this equation, the first term is the total mass ejected by the flow that has been incorporated
into the WS and the second term is the actual mass of the WS. These two terms increase with time but their
difference remains finite because the shocks will weaken with time.

One can define the fraction of mass lost by radiation as
\be
\Delta m = \frac{\int_0^t{L_r(t) c^{-2}} dt}{\int_{\tau_1}^{\tau_2} \dot m(\tau) \gamma(\tau) d \tau}
= 1 - \frac{ \mws \gammaws }{\int_{\tau_1}^{\tau_2} \dot m(\tau) \gamma(\tau) d \tau}.
\ee
This fraction $\Delta m \rightarrow 0$ when $t \rightarrow \infty $
because the total mass lost by radiation in the LHS of equation (\ref{mass_lost}) is finite. 
Thus, one has to evaluate $\Delta m$ at a
finite time to determine the importance of radiation losses in the dynamics of the working surface.

The total momentum lost by radiation $\int_0^t L_r(t) c^{-2} \betaws dt$ can be obtained
from equation ({\ref{mom_eq}}). In particular, in the UR regime where $\betaws \sim 1$,  
the mass and momentum losses are the same. In this regime, $\Delta m$ measures the importance
of both momentum and mass losses.

If $\Delta m \ll 1$, radiation losses will not change the relativistic mass of the WS
significantly. 
 Note that in the constant velocity phase, radiation losses, which change
the relativistic mass of the WS according to equation (\ref{energy_eq}), 
do not affect the velocity $\betaws$ because the LHS of eq. (\ref{princ_eq}) is zero (since $d \betaws/dt =0$). 
Nevertheless, these losses do change the momentum of the WS at the critical time and the dynamics of the WS 
in the second decelerating/accelerating phase of Case I and II, respectively.

We choose to evaluate $\Delta m$ at the critical times (eqs. [\ref{tcI}] and [\ref{tcII}]), which correspond to the end
of the constant velocity phase. For Case I, $\tau_{2c} = \Delta \tau$, thus,
$
\int_{\tau_{1c}}^{\Delta \tau} \dot m(\tau) \gamma(\tau) d \tau = \dot m_1 \gamma_1 \Delta \tau a \lambda
(1+ \lambda);
$
for Case II,  $\tau_{1c} = -\Delta \tau$, thus,
$
\int_{-\Delta \tau}^{\tau_{2c}} \dot m(\tau) \gamma(\tau) d \tau = \dot m_1 \gamma_1 \Delta \tau
(1+ \lambda).
$
Then, using equations (\ref{mwsI}), (\ref{mwsII}), one can show that in both cases I and II,
the mass fraction is
\be
\Delta m = 1 - \frac{\gammaws}{\gamma_1} \frac{\left ( b + a \lambda \right)}{a \lambda \left(1+\lambda\right)} .
\ee
In the UR regime, where $a\sim 1$, $\lambda \sim (b r)^{1/2}$, and $\gammaws$ is 
given by equation (\ref{gammaws0}), this expression reduces to
\be
\Delta m = 1 - \frac{r\left(b+\lambda \right)}{\lambda \left(\lambda + r^2\right)^{1/2} \left(1+\lambda\right)^{1/2}},
\ee
which is a function only of the gamma ratio, $r=\gamma_2/\gamma_1$, and the mass-loss rate
ratio, $b = \dot m_2 / \dot m_1$. 

\begin{figure}
\includegraphics[width=84mm]{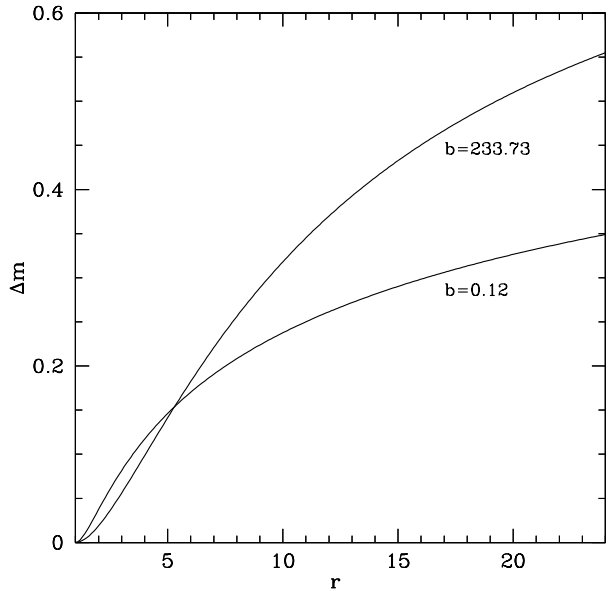}
\caption{Fraction of mass lost by radiation as a function of the gamma ratio $r=\gamma_2/\gamma_1$.
Each curve is labeled by the different values of the mass-loss rate ratio $b = \dot m_2/\dot m_1$.}

\label{Deltam}
\end{figure}

Figure \ref{Deltam} shows $\Delta m$ as a function of $r$ for 
the $b$ values of the models presented in Table 1. One can see that the mass fraction 
is $\Delta m \sim 7$\% for the model of GRB 070318, 
but it reaches values $\Delta m > 50$\% for the model of GRB 080413B. In the last case,  
  radiation losses clearly affect the dynamics of the WS: mass and momentum are not conserved.

\section{Conclusions}

We have developed a new formalism that describes the dynamics of an internal WS in a relativistic
jet produced by variations in the source injection velocity. The WS is formed when a fast flow overtakes
a previous slower flow. This formalism takes into account that 
the momentum is not conserved because relativistic mass is lost by radiation, 
in contrast with non relativistic flows.

Assuming step function variations of the injection velocity and mass-loss
rate we find analytic solutions for the WS velocity and luminosity. We consider two cases: when a pulse of
fast material reaches the slow downstream wind (Case I); and when a pulse of slow 
material is pushed by fast upstream wind (Case II). 
In the initial phase, the WS is bounded by 2 shocks: 
one shock incorporates the material from the fast (slow) pulse, the other shock
incorporates material from the slow (fast) wind. In this phase, the velocity of the WS is constant. When the
material from the fast (slow) pulse is completely incorporated into the WS, only one shock remains: in 
Case I, the WS is decelerated as more mass is added to the shock by the slow downstream wind; 
in Case II, the WS accelerates, pushed by the fast upstream wind. The WS luminosity in the constant velocity
phase is constant, and decreases with time when one of the shocks disappears. To apply these models to 
observed GRBs we assume that a constant fraction $\epsilon$ of this energy is emitted in gamma rays.
  
In the UR limit, the ratio of the Lorentz factors $r=\gamma_2/\gamma_1$ and the mass-loss rates $b=m_2/m_1$ of the 
relativistic flows that collide can be obtained directly by fitting the light curves of GRBs. 
 As an example, we fit the light curves
of the GRBs 080413B and 070318 with the Case I and Case II models, respectively. 
For GRB 080413B we obtain the ratios $\gamma_2/\gamma_1 = 22.5$ and $\dot m_2/\dot m_1 = 233.7$;   
for GRB 070318  the fit gives lower ratios,  $\gamma_2/\gamma_1 = 2.6$ and $\dot m_2/\dot m_1 = 0.1$. 
Since the WS is moving towards the observer at relativistic speeds ($\gamma_{\rm ws0} \sim 100 - 800$), 
one has to correct the  observed gamma ray fluxes
for Doppler boosting. Assuming $\gamma_1=100$,
we estimate mass-loss rates of the jets  between
$\dot m_1 \sim 10^{-5}  - 1 \, {\rm M}_\odot \, {\rm yr}^{-1}$.
Note that the jet kinetic power is $P_{kin} = \gamma \dot m c^2 \sim 10^{44-48} {\rm erg s^{-1}}$. This is much smaller than
the associated isotropic luminosity, $L_{iso} = 4 \pi D^2 F_{GRB} /\epsilon \sim 10^{53-54} {\rm erg s^{-1}}$, uncorrected for relativistic effects. 
In fact,  the isotropic luminosity has no
physical meaning for our relativistic jet models, which are able to produce the Doppler boosted gamma
ray flux observed at the Earth.

 We also evaluate the fraction of the injected mass lost by radiation.  
For the model of the source GRB 070318 this fraction is $\sim$ 7\%, while 
for the source GRB 080413B, one finds that more than 50\% of
the injected mass is lost by radiation. 
Therefore, in the latter source, radiation losses change significantly the relativistic mass
of the WS and affect its dynamics.

The step function variability of the source injection velocity and
mass-loss rate is a simple approximation to the real time variability  
of the injection parameters. 
Other functional time variations of these parameters can be easily 
 implemented in our formalism by integrating the equations in \S 3 numerically. 
Nevertheless, in the UR regime our analytic 
model is very useful to determine the ratios of important physical 
parameters (the gamma ratio and the mass-loss rate ratio)
without introducing any further assumptions.

In a future work we will study the high energy emission produced by these internal shocks, in particular,
the fraction of energy emitted as gamma rays.

\section*{Acknowledgments}
J.C., S.L., M.F., R.F.G., and A.H. were supported by PAPIIT-UNAM IN100412  and IN100511. 
J. C. was also supported by CONACyT 61547. We  thank Jose Ignacio Cabrera for 
providing us with the observational data and Anabella Araudo for useful suggestions.
 We also thank an anonymous referee for useful comments that helped to improve the paper.



\appendix

\section[]{Collision of relativistic particles}
\label{App_A}

We discuss a simple example of an inelastic collision of 3 relativistic particles that
shows that when energy is radiated in a shock, momentum is not conserved because
the relativistic mass, that includes the Lorentz factor $\gamma_i$ of internal motions, is not conserved.

Consider two particles in the laboratory 
with masses $m_k$, velocity $\beta_k$, momentum $P_k = m_k \gamma_k c \beta_k$,
and energy, $E_k = m_k \gamma_k c^2$,  for particles $k=1,2$. We consider the collision
between these two particles, where $\beta_2 > \beta_1$. In the laboratory reference frame 
the momentum conservation (divided by $c$) gives, 
\be
m_1 \gamma_1 \beta_1 +m_2 \gamma_2 \beta_2 = (m_1 + m_2) \gamma_{i12} \gamma_{12} \beta_{12},
\label{pc}
\ee
where $\gamma_{i12}$ is the Lorentz factor that 
corresponds to internal motions, and $\gamma_{12}$ corresponds to the
bulk motion of the new particle 1-2  with velocity $\beta_{12}$.
Energy conservation (divided by $c^2$) gives
\be
m_1 \gamma_1 + m_2 \gamma_2 = (m_1 + m_2) \gamma_{i12}\gamma_{12}. 
\label{ec}
\ee
The ratio of these two equations gives the velocity of particle 1-2,
\be
\beta_{12} = {m_1 \gamma_1 \beta_1 + m_2 \gamma_2 \beta_2 \over m_1 \gamma_1  + m_2 \gamma_2 },
\label{beta12}
\ee
and the Lorentz factor for internal motions is obtained from equation (\ref{ec})
\be
\gamma_{i12} = {m_1 \gamma_1  + m_2 \gamma_2 \over (m_1+m_2)  \gamma_{12}}.
\label{gammai12}
\ee

Consider a momentum center reference system where the total momentum is zero. This system moves with respect to the
laboratory frame with the speed
$\beta_{12}$ defined by equation (\ref{beta12}). The Lorentz transformation of the energy
 of the two particles before collision in this primed system gives
\begin{eqnarray}
E_k^\prime=\gamma_{12} \left({m_k \gamma_k c^2 - c \beta_{12}
 (m_k \gamma_k c \beta_k ) }\right)
\nonumber
\end{eqnarray}
\be
= m_k \gamma_k \gamma_{12} \left ( 1 - \beta_k \beta_{12} \right) c^2,
\ee
and the momentum transformation gives
\be
P_k^\prime = m_k \gamma_k \gamma_{12} c \left(\beta_k - \beta_{12} \right).
\ee
Using equation (\ref{beta12}), one can check that $P_1^\prime + P_2^\prime = 0.$ 

In this system, after the collision, the
new particle 1-2 will be at rest. Before any radiation is emitted,  the initial energy before collision $E^\prime_{initial}$ is conserved;
thus, after collision  it will become the internal energy of the
particle 1-2, $E^\prime_{internal}$,
\begin{eqnarray}
E^\prime_{initial}=E_1^\prime+E_2^\prime = {\left( m_1 \gamma_1 + m_2 \gamma_2 \right) c^2
 \over \gamma_{12}}
\nonumber
\end{eqnarray}
\be
= \left(m_1+m_2 \right) \gamma_{i12}^\prime c^2 = E^\prime_{internal},
\label{eprime}
\ee
where we used equation (\ref{beta12}) to calculate $E^\prime$, and $\gamma_{i12}^\prime$ is associated with the internal
motions of particle 1-2 in the momentum center reference system.  Solving this equation for $\gamma_{i12}^\prime$, one gets
\be
\gamma_{i12}^\prime = {m_1 \gamma_1  + m_2 \gamma_2 \over (m_1+m_2)  \gamma_{12}}
= \gamma_{i12},
\ee
where we used equation (\ref{gammai12}); i.e., the Lorentz factor associated with internal motions is the same in
both reference systems. Thus, after collision the new particle 1-2 has a larger
mass, $(m_1+m_2)\gamma_{i12}$.

Now, let us assume that the particle 1-2 radiates isotropically (in the momentum center system) 
all the available energy, thus, the momentum of particle 1-2 will not change, $P_{12}^\prime=0$.
The maximum energy that can be radiated by particle 1-2 is
\begin{eqnarray}
E_r^\prime = E_{internal}^\prime - E_{0}^\prime
\nonumber
\end{eqnarray}
\be
= {c^2 \over \gamma_{12}} \left[m_1 \gamma_1 + m_2 \gamma_2
- (m_1 + m_2) \gamma_{12} \right],
\label{Rad}
\ee
where $E_{0}^\prime = (m_1+m_2) c^2$ is the rest energy of particle 1-2, and we used equation(\ref{eprime}).

In the laboratory reference frame, this particle 1-2 that has radiated all the available energy now has
an energy and momentum given by the Lorentz transformations
\be
E_0 = \gamma_{12}\left( E_0^\prime + c \beta_{12} P_{12}^\prime \right) = (m_1+m_2)\gamma_{12} c^2;
\label{E0}
\ee
\be
P_0 = \gamma_{12} \left( P_{12}^\prime + {\beta_{12} \over c} E_0^\prime \right) 
= (m_1+m_2) \gamma_{12} c \beta_{12}.
\label{P0}
\ee
Comparing these equations with the RHS of 
equations (\ref{pc}) and (\ref{ec}) one can see that after radiation, particle 1-2 has lost
energy and momentum.

On the other hand, the velocity of particle 1-2 after radiation in the laboratory reference frame is given by
\be
\beta\equiv  {c P_0 \over E_0} = \beta_{12}.
\label{betafinal}
\ee
Comparing with equation (\ref{beta12}) one can see that, after radiating all the available energy, particle 1-2
conserves its velocity.

Now, let us consider the collision of three particles with rest masses $m_k$ and
velocities $\beta_k$, respectively, for $k=1,2,3$.
We want to obtain the velocity of the particle that results from the collision of the three particles.

As discussed above, $\beta_2 > \beta_1$, such that particle 2 will collide with particle 1. The resulting velocity of
the combined particle 1-2 and internal Lorentz factor are given by equations 
(\ref{beta12}) and  (\ref{gammai12}). Now we consider
the collision of particle 1-2 with particle 3 that has $\beta_3 > \beta_{12}$.

If the new particle 1-2 collides with particle 3 before it radiates away energy,
the conservation of momentum gives
\begin{eqnarray}
(m_1+m_2) \gamma_{i12} \gamma_2 \beta_{12} + m_3 \gamma_3 \beta_3 =
\nonumber
\end{eqnarray}
\be
=(m_1 + m_2 + m_3) \gamma_{i123} \gamma_{123} \beta_{123}\,,
\label{pcons3}
\ee
and the conservation of energy gives
\be
(m_1+m_2) \gamma_{i12} \gamma_2  + m_3 \gamma_3 
=(m_1 + m_2 + m_3) \gamma_{i123} .
\label{econs3}
\ee
In this case, where no energy is radiated before the collision of the 3 particles, 
the velocity of the new particle 1-2-3 is given by the ratio of equations (\ref{pcons3}) and (\ref{econs3}) 
\begin{eqnarray}
\beta_{123}^{ ad}= {(m_1+m_2) \gamma_{i12} \gamma_2 \beta_{12} + m_3 \gamma_3 \beta_3
\over (m_1+m_2) \gamma_{i12} \gamma_2  + m_3 \gamma_3 }
\nonumber
\end{eqnarray}
\be
= {m_1 \gamma_1 \beta_1 + m_2\gamma_2 \beta_2 + m_3 \gamma_3 \beta_3
\over m_1 \gamma_1 + m_2 \gamma_2 + m_3 \gamma_3},
\label{beta123ad}
\ee
where we used equation (\ref{pc}) to obtain the last equality. Note that the RHS of the above equation is also
obtained from the conservation of momentum and  energy of the collision of the 3 particles. 
If the new particle 1-2-3 radiates all its internal energy,  as shown in equation (\ref{betafinal}),
 it will conserve this velocity.

We now consider the case when particle 1-2 radiates all the available energy before colliding with
particle 3. In this case, particle 1-2 has the momentum and
the energy given by equations (\ref{P0}) and (\ref{E0}). Then, the equation of momentum conservation is
\be
(m_1+m_2) \gamma_2 \beta_{12} + m_3 \gamma_3 \beta_3
=(m_1 + m_2 + m_3) \gamma_{i123} \gamma_{123} \beta_{123},
\label{pcons3n}
\ee 
and the conservation of energy is
\be
(m_1+m_2) \gamma_2  + m_3 \gamma_3 
=(m_1 + m_2 + m_3) \gamma_{i123}  \gamma_{123}.
\label{econs3n}
\ee
In this case, when energy is radiated by particle 1-2 before collision with particle 3, 
the velocity of the new particle 1-2-3 is given by the ratio of equations (\ref{pcons3n}) and (\ref{econs3n}), 
\be
\beta_{123}^{rad}= {(m_1+m_2) \gamma_2 \beta_{12} + m_3 \gamma_3 \beta_3
\over (m_1+m_2) \gamma_2  + m_3 \gamma_3 }.
\ee
This velocity is different from the velocity of the adiabatic case, equation (\ref{beta123ad}).
Therefore, the collision of particles that radiate their internal energy does not conserve
momentum! This happens because when energy is radiated before collision by particle 1-2 above, its momentum
decreases because its relativistic mass $(m_1+m_2)\gamma_{i12}$ decreases, i.e.,  $\gamma_{i12} \rightarrow 1$.

Therefore,  when energy is radiated in a shock in the 
collision of 3 relativistic particles, one cannot
obtain the velocity of the shock by assuming momentum conservation.

\begin{figure}
\includegraphics[width=84mm]{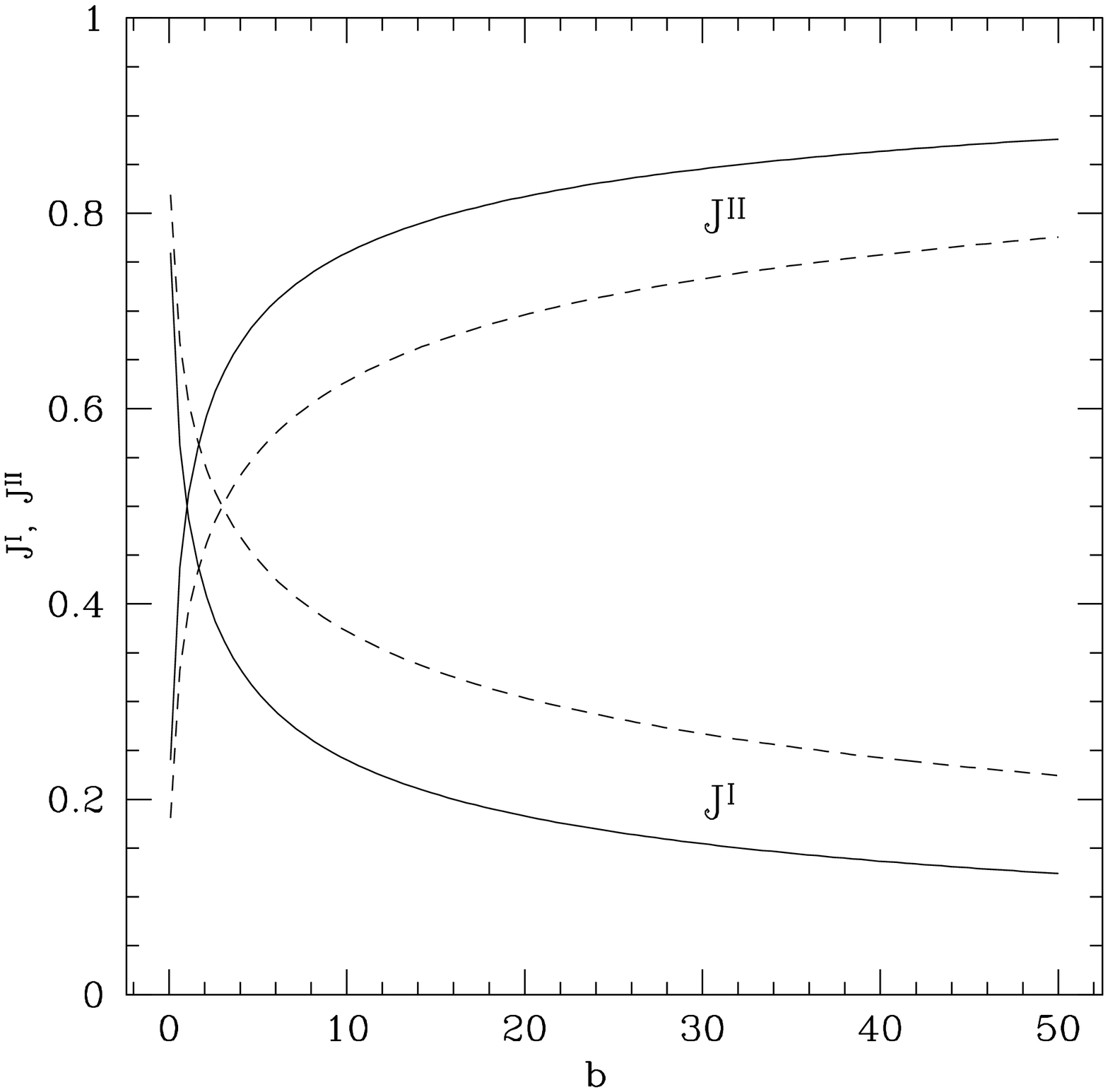}
\caption{Luminosity jumps $J^I$ and $J^{II}$ as function of the mass-loss rate ratio
$b$ for two $\gamma$ ratios: $r \rightarrow 1$ (solid lines) and $r=3$
(dashed lines).}
\label{J_b}
\end{figure}

\section[]{The Ultra-relativistic Case}
\label{App_B}

Here we will consider the ultra-relativistic (UR) limit, $\beta \simeq 1$, where the Lorentz
factor $\gamma \gg 1$. In this limit, $a \sim 1$,  and 
one can expand  $\beta \simeq 1 - 1/(2\gamma^2)$. Thus, 
$\beta_{\rm ws} -
\beta_1 \simeq (\gamma_{\rm ws}^2 - \gamma_1^2)/2\gamma_1^2 
\gamma_{\rm ws}^2$, and $a-1 \sim (\gamma_2^2-\gamma_1^2)/(2 \gamma_1^2 \gamma_2^2)$.

We define the 
normalized Lorentz factors $r_1 = \gammaws/\gamma_1$ and $r_2=\gammaws/\gamma_2$. 

In the constant velocity phase, the velocity  of the WS, given by 
equation (\ref{beta_first}), can be written as
\be
\beta_{\rm ws0} 
\simeq 1 - {{\gamma_2^2 + \lambda \gamma_1^2}\over{2 \gamma_1^2 \gamma_2^2
 (1 + \lambda)}}\,,
\ee
where $\lambda \sim  (br)^{1/2}$ (eq. [\ref{lambda}] with $a= 1$).
Thus, the constant Lorentz factor is
\be
\gamma_{\rm ws0} \simeq {{{(1 + \lambda)}^{1/2} \gamma_1 \gamma_2}
\over{{(\gamma_2^2 + \lambda \gamma_1^2)}^{1/2}}}\,,
\label{gammaws0}
\ee
and the normalized Lorentz factors are constant and are given by
\be
r_{1}^0= {\left(1 + \lambda \right)^{1/2} r \over \left( r^2 + \lambda \right)^{1/2} }\,,
\quad {\rm and} \quad
r_{2}^0={r_{1}^0 \over r}.
\label{r120}
\ee

In the decelerating Case I and the accelerating Case II, the normalized Lorentz factors depend on
time and are simply given by (eqs. [\ref{betawstau1}] and [\ref{betawstau2}])
\be
r_{1}^I = {\T^I+2 \over \T^I-2}, \quad {\rm and} \quad r_{2}^{II} = {\T^{II}-2 \over \T^{II}+2} ,
\label{r12t}
\ee
where the time functions $\T^i(\tilde t)$ are given by equations (\ref{calT1_t}) and (\ref{calT2_t}), respectively.

Thus, the luminosity in the constant velocity phase (eq. [\ref{L_0}]) is
\begin{eqnarray}
\tilde L_{UR}^0 & \simeq & 
{\gamma_{ws0}^2- \gamma_1^2  \over 2 \gamma_1^2 \gamma_{ws0}^2} 
 \left[ \lambda {\left(\gamma_2 - \gamma_{ws0}\right) \over \gamma_2} - {\left(\gamma_{ws0} - \gamma_1 \right) \over \gamma_1} \right], \nonumber \\
&\simeq &
 {1 \over 2 \gamma_1^2}  
 {\left( r_{1}^0 + 1\right)\left( r_{1}^0 - 1\right)^2  
 \left(r_{1}^0-r_{2}^0\right) \over (r_{1}^0)^2\left(1+r_{2}^0\right)},
 \label{LUR0}
\end{eqnarray}
where the last term was obtained by solving for $\lambda$ in equation (\ref{r120}).
 The luminosity of the decelerating Case I (eq. [\ref{L_I}]) in the UR case simplifies to
\be
\tilde L_{UR}^I (\tilde t)\simeq
{1 \over 4 \gamma_1^2} { \left(r_{1}^I+1\right) \left(r_{1}^I -1 \right)^3\over  (r_{1}^I)^2  } 
=  {32  \over  \gamma_1^2} {\T^I \over \left([\T^I]^2-4\right)^2} ,
\label{LURI}
\ee
and the luminosity of the accelerating Case II (eq. [\ref{L_II}]) simplifies to
\begin{eqnarray}
\tilde L_{UR}^{II}(\tilde t) &\simeq&
{\lambda^2  \over 4 \gamma_2^2} { \left(r_{2}^{II}+1\right) \left(1- r_{2}^{II} \right)^3\over  (r_{2}^{II})^2  } 
\nonumber \\
& = & {32 \lambda^2 \over \gamma_2^2} {\T^{II} \over \left([\T^{II}]^2-4\right)^2} .
\label{LURII}
\end{eqnarray}
Finally, the dimensional luminosities are
 \be
 L^i_{UR} = l^i {32 \T^i(t) \over \left([\T^i(t)]^2 - 4 \right)^2},
 \label{LGRBUR}
\ee
where
$l^{I}= \epsilon c^2  \dot m_1 /  \gamma_1$ for Case I, 
and  $l^{II} = \epsilon c^2 \dot m_2 / \gamma_2$
for Case II.

\begin{figure}
\includegraphics[width=84mm]{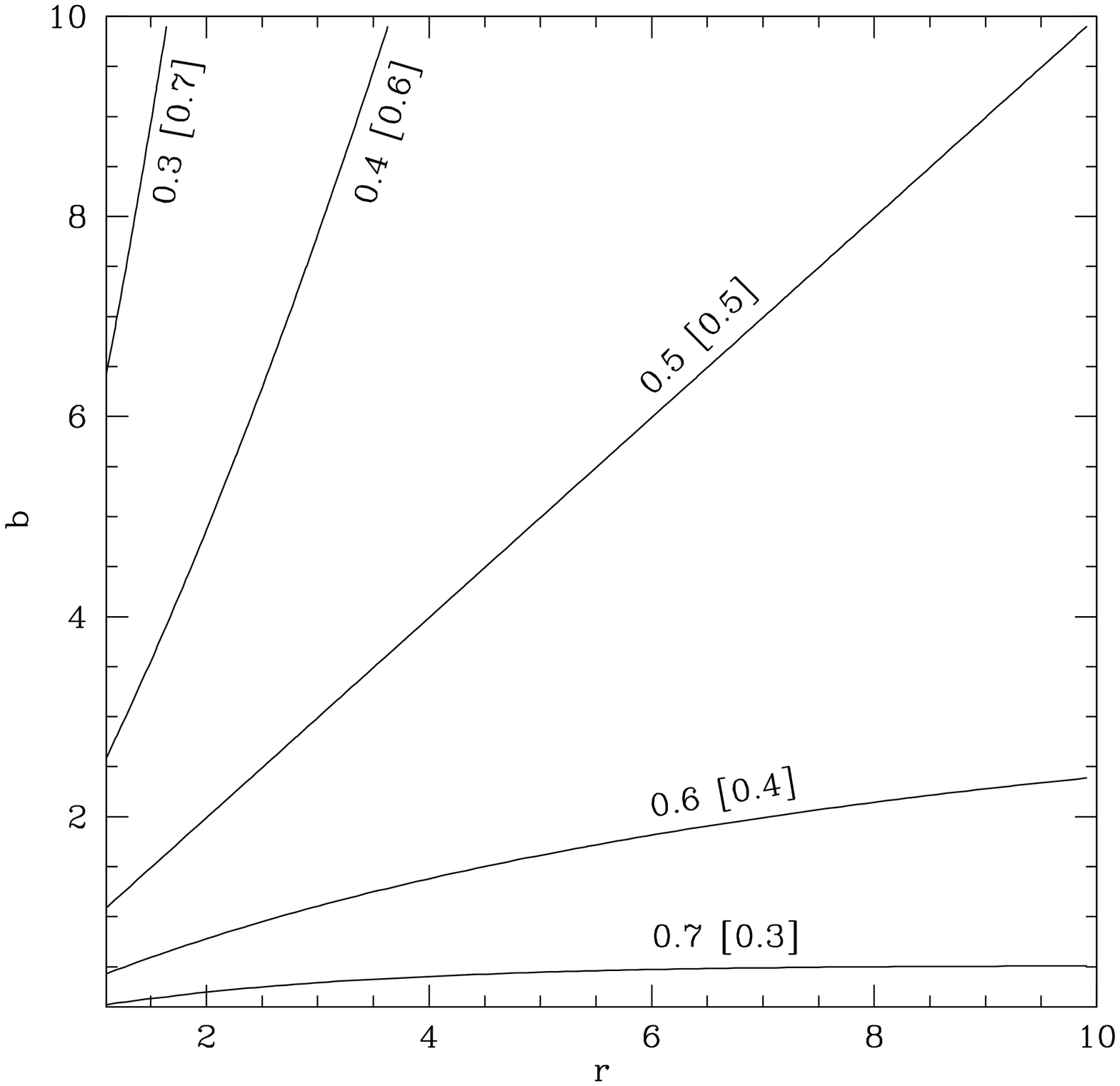}
\caption{Isocontours of the luminosity jump as function of the mass-loss rate ratio, $b$, and the gamma ratio, $r$. Each contour is
labeled by the value of the jump for Case I and, in square brackets, the value of the jump for Case II.}
\label{b_r}
\end{figure}

We now examine the  ``jump'' in the luminosity at the critical time $t_{c}^i$. At this time,
$r_{1,I}$ and $r_{2,II}$ coincide with $r_{1,0}$ and $r_{2,0}$, respectively, given by eqs. (\ref{r120}), because the velocity
of the WS is a continuous function of time.
Then, we define the jump of the luminosities at $t_{c}^i$ as
\begin{eqnarray}
J^I =1- { L_{UR}^I ( t_{c}^I) \over L_{UR}^0} 
 ={ (r_{1}^0 +1)( 1-r_{2}^0) \over 2 (r_{1}^0-r_{2}^0) }  ,
\label{LURI0}
\end{eqnarray}
for Case I, and
\begin{eqnarray}
 J^{II}= 1- { L_{UR}^{II} ( t_{c}^{II}) \over  L_{UR}^0} =
 {\left(r_{1}^0 - 1\right) \left(1+r_{2}^0\right)\over 2 \left(r_{1}^0- r_{2}^0\right)} ,
\label{LURII0}
\end{eqnarray}
for Case II, where we used equations (\ref{LUR0}), (\ref{LURI}), and (\ref{LURII}).
The luminosity jumps  are shown in Figure \ref{J_b} as a function of the
mass-loss rate ratio $b$ for two values of the gamma ratio  $r$: $r \rightarrow 1$
(solid line) and  $r=3$ (dashed line).  One can see that small luminosity
jumps are predicted for large values of $b$ in Case I; the opposite is true
for Case II.
The figure also shows that the summ $J^I + J^{II} = 1$  because at the
critical time, the WS of Case I and Case II are each one bounded by only
one of the shocks that provide the total luminosity, $L^0$, in the constant
velocity phase. This fact is also presented in Figure \ref{b_r} by
isocontours of the luminosity jumps $J^I$ and $J^{II}$ as functions of
the mass-loss rate ratio $b$ and gamma ratio $r$. 
\footnote{It can be shown from equations (\ref{solpm}), (\ref{L_0}),
(\ref{L_I}), and (\ref{L_II}), that $L^0=L^I+L^{II}$, is a general result, not only valid for the UR case.}

\begin{table}
 \centering
 \begin{minipage}{165mm}
  \caption{Coefficients}
  \begin{tabular}{@{}cccccc@{}}
\hline
\hline
Case & $a_0$ & $a_1$ & $a_2$ & $b_1$ & $b_2$ \\ 
\hline
I  & -2.0000  & -1.6724 &  -0.3646 & 0.7635 & 0.2431 \\
II  & -0.6667 & -0.5494 &  -0.2651 & 0.6247 & 0.1768 \\ 
\hline
\end{tabular}
\end{minipage}
\end{table}

\section{The Critical Time Function, $\T_{c}$}
\label{alpha_fit}

This Appendix describes a procedure to obtain the critical time function, $\T_c$, from a fit of the GRB light curve.
Let us assume that the constant velocity phase starts at an observed time $t_0$, then, 
the decaying wing of the GRB light curve starts at $t_e=t_0+t_c$,
where $t_c$ is the critical time ($t_e$ measures the end of the constant velocity phase).
We further assume that the wing of the GRB light curve can be fit by a power-law,
$L = B (t-t_0)^\alpha$, $t>t_e$. Consider three times: $t_n=n t_{c}+t_0$, $n=1,2,3$, 
and the corresponding dimensional luminosities, $L (\T_n(t_n))$, which are given by equation (\ref{LGRBUR}). 
Then, one can
construct the relation
\be
n^\alpha = {\T_n  \left( \T_{c}^2 - 4 \right)^2 \over  \T_{c}  \left(\T_n^2 - 4 \right)^2 } ,
\ee
which can be solved for the time function $\T_n$ as
\be
\T_n^4 - 8 \T_n^2 - { \left( \T_{c}^2 - 4 \right)^2 \over n^\alpha \T_{c} } \T_n+ 16 =0, \quad n=2,3.
\label{Tn}
\ee
 These are two quartic equations for time functions $\T_n (\T_{c}, \alpha)$.

\begin{figure}
\includegraphics[width=84mm]{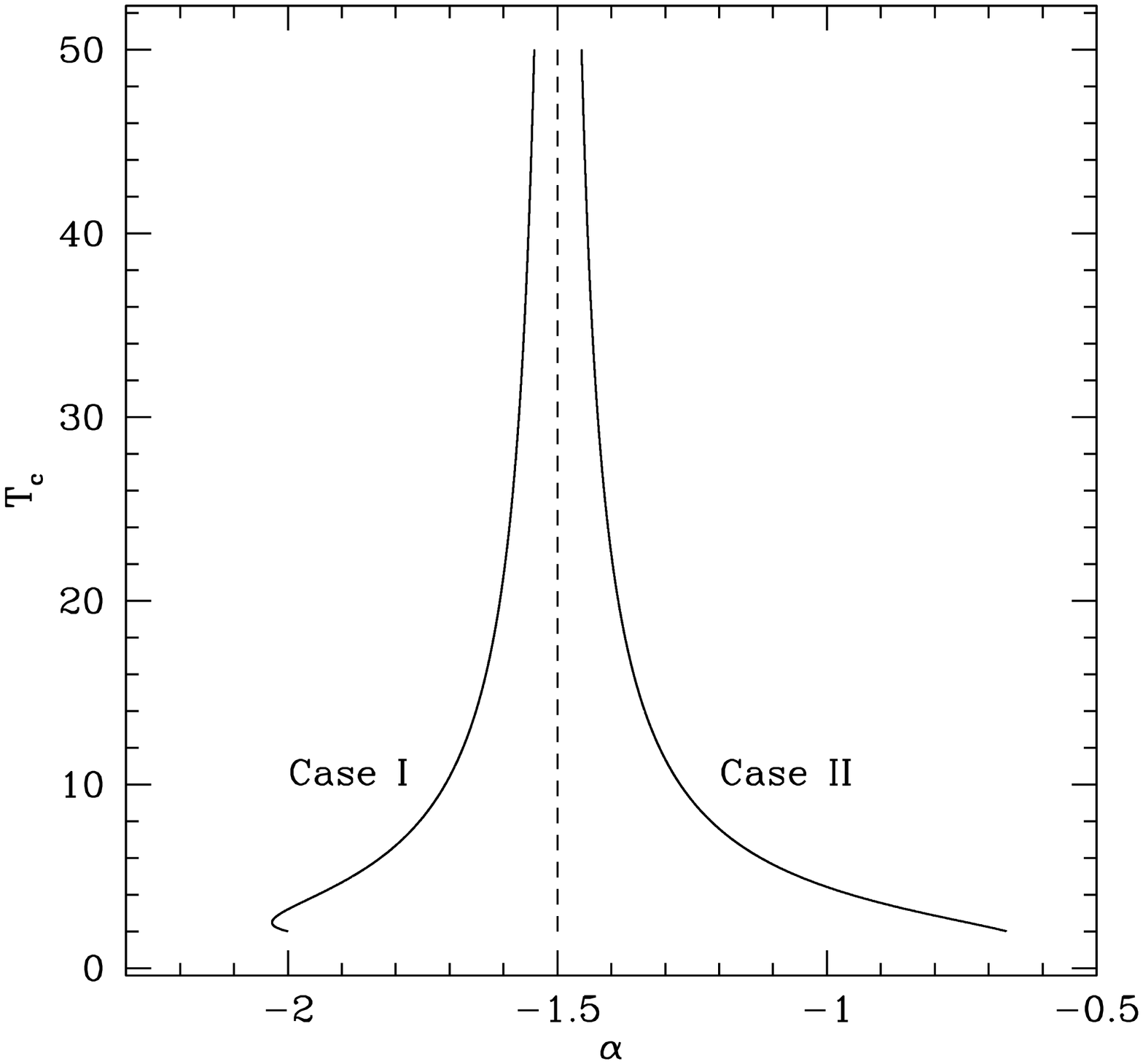}
\caption{Relation between the exponent $\alpha$ and the critical time, $\T_c$, 
for Case I and Case II, obtained from equation (\ref{alphaTeq}).}
\label{alphaT}
\end{figure}

 Equations (\ref{calT1_t}) and (\ref{calT2_t}) give the relation between the time $t$ and $\T$ for Cases I and II,
 respectively. Let us define
 $\mu^I(\T) = \T^2/2 + 4 \T + 4 \ln\T$, for Case I, and 
$\mu^{II}(\T) = \T^2/2 - 4 \T + 4 \ln\T$, for Case II.
 Then, one can construct the relation
\be
{\cal H^{I,II}} = \mu^{I,II}(\T_3) - 2 \mu^{I,II}(\T_2) + \mu^{I,II}(\T_{c}) = 0 .
\label{HI_II}
\ee
When one substitutes $\T_n (\T_{c}, \alpha)$ from equation (\ref{Tn}) in this equation, one
obtains a relation ${\cal H^{I,II}}(\T_{c}, \alpha)=0$. This relation gives two distinct curves for the critical time
function $\T_c(\alpha)$ for Case I and II.
These curves can be fit
by the
Pad\'e polinomials
\be
\alpha = { a_0+a_1 \left(\T_{c} - 2\right) + a_2 \left(\T_{c} -2 \right)^2  \over
1+b_1 \left(\T_{c} - 2\right) + b_2 \left(\T_{c} -2 \right)^2 },
\label{alphaTeq}
\ee
where the coefficients are given in Table 2. These curves are shown in 
Figure \ref{alphaT}. In particular,  the values of the exponent
$\alpha$ are restricted to the interval $[-2.02662,-3/2]$ for Case I, and 
$[-3/2, -2/3]$ for Case II. Thus, from the slope $\alpha$, one can determine
to which case corresponds the observed wing of the GRB light curve: Case I (decelerating WS) vs Case
II (accelerating WS).

\section{Determination of Fundamental Parameters: gamma ratio and
mass-loss rate ratio in the Ultra relativistic Case.}
\label{fund_par}

The gamma ratio and mass-loss rate ratio can be obtained by the following procedure.

I) Given the GRB light curve, one chooses the time for the beginning of the velocity pulse, $t_0$ (this corresponds
to $t=0$ in the models), and the time for the end of the constant velocity
phase, $t_{e}=t_0+t_c$,  where the critical time is $t_c$.

II) From a numerical fit of the GRB light curve wing,
$L=B(t-t_c)^\alpha$, the critical time $\T_{c}$ for Case I or Case II 
can be obtained  from equation (\ref{HI_II}). 

III) From the observations, the jump $J$ of the luminosities at $t_c$, 
defined by equation (\ref{LURI0}) and equation (\ref{LURII0}) for Case I and
Case II, respectively, are estimated. Also, from these equations, writing
the gamma ratio $r= r_{1}^0/r_{2}^0$, with $r_{1}^0={(\T_{c}+2)/(\T_{c} -2)}$
for Case I and $ r_{2}^0= {(\T_{c}-2)/(\T_{c} +2)}$ for Case II
(see eq. [\ref{r12t}]), it can be shown that, 
\be
r={    { \left(\T_{c}+2\right) \left[J \left(\T_{c}-2 \right) -\T_{c} \right] } \over
{\left( \T_{c}- 2 \right) \left[ J \left( \T_{c} + 2 \right) - \T_{c} \right]}      }\,,
\ee
where this equation applies for both cases I and II.

IV) Finally, the parameter $\lambda = b^{1/2} r^{1/2}$ is calculated
from equation (\ref{r120}), that is,

\begin{eqnarray}
\lambda_I= { \left[ \T_c - J \left( \T_c -2 \right) \right]^2  \over
J \left(1-J \right) \left( \T_c-2 \right)^2}\,\quad , \,\mbox{and}
\end{eqnarray}

\be 
\lambda_{II}= { J \left(1-J\right) \left(\T_c+2 \right)^2 \over
\left[ \T_c - J \left(\T_c + 2\right) \right]^2 }\,.
\ee

Therefore, only the ratios $\lambda$ and $r$ can be obtained independently
by fitting the observations in the UR case. Finally, from $\lambda$ and
$r$ one can obtain the mass-loss rate ratio $b=\lambda^2/r$.

\section{Doppler boosting}
\label{Dopp_boost}

As found by Lind and Blandford (1985),  for a WS in a relativistic jet approaching the observer with a velocity $\betaws$,
the observed flux at a given frequency, 
$F_\nu$ is increased with respect to an emitter at rest by a factor $\delta_{\rm ws}^3$, where 
the Doppler factor is $\delta_{\rm ws} = 1/[\gammaws(1-\betaws \cos\theta)]$, where $\theta$ is the angle
between the observer and the jet axis.

In particular, for an optically thin WS, the observed flux is given by
\be
F_\nu = {\delta_{\rm ws}^3 \over D^2} \int_{V^\prime} j^\prime_{\nu^\prime} dV^\prime,
\ee
where $D$ is the distance to the jet source and
$j^\prime_{\nu^\prime}$ is the volume emissivity (${\rm erg\, cm^{-3} s^{-1} str^{-1} Hz^{-1}}$) 
at frequency $\nu^\prime = \nu/\delta_{\rm ws}$ measured in the rest frame of the emitting source, and the integration is carried
over the volume $V^\prime$.

Integrating the flux over a $4 \pi$ solid angle and over the frequency $\nu^\prime$,
one gets
\begin{eqnarray}
4 \pi \int_0^\infty F_\nu d \nu^\prime 
= {\delta_{\rm ws}^3 \over D^2} 4 \pi \int_0^\infty \int_{V^\prime}  j^\prime_{\nu^\prime} dV^\prime d \nu^\prime.
\nonumber
\end{eqnarray}
Given the transformation $\nu^\prime = \nu/\delta_{\rm ws}$, the integral in the LHS of the above equation is
\be
\int_0^\infty F_\nu d \nu^\prime = {1 \over \delta_{\rm ws}} \int_0^\infty F_\nu d \nu,
\ee
therefore, the bolometric flux is 
\be
F_{\rm bol} \equiv \int_0^\infty F_\nu d \nu  = {\delta_{\rm ws}^4 \over 4 \pi D^2} L^\prime,
\ee
where $L^\prime$ (${\rm erg \, s^{-1}}$) is the bolometric luminosity or the total emitted power of the source. 

In our model, the luminosity of the WS in equation (\ref{lum}), $L_r$, is measured in the rest frame of the source of the jet.
This total emitted power is a Lorentz invariant for any source that emits isotropically in its instantaneous frame of
reference (Rybicki \& Lightman 1979). Thus,
$L_r= L^\prime$.
Furthermore, since we consider that a fraction $\epsilon$ 
of the bolometric luminosity is emitted in gamma rays,  $L_{GRB} = \epsilon L_r$ (see \S 4), 
then, the observed flux in gamma rays,
$F_{GRB} = \epsilon F_{\rm bol}$, is given by
\be
F_{GRB} = {\delta_{\rm ws}^4 \over 4 \pi D^2} L_{GRB}.
\ee
This equation can also be written in terms of the non dimensional luminosity in equation (\ref{LGRBnondim}) as
\be
F_{GRB} =  {\delta_{\rm ws} ^4 \over 4 \pi D^2} \tilde L \, \epsilon \, \dot m_1 \gamma_1 c^2.
\ee

The case of an optically thick WS, discussed by Lind \& Blandford (1985), follows in a straight forward fashion. One also obtains the
$\delta_{\rm ws}^4$ factor between the emitted and observed bolometric fluxes because the extra $\delta_{\rm ws}$ factor
is due to the integration in frequency.

\bsp

\label{lastpage}

\end{document}